# Tissue engineering for periodontal ligament regeneration: Biomechanical specifications

R.Gauthier[1,2], C.Jeannin[1,2,3], N.Attik[1,2], A-M Trunfio-Sfarghiu[4], K.Gritsch[1,2,3], B.Grosgogeat[1,2,3]

[1] Univ Lyon - Claude Bernard Lyon 1, UMR CNRS 5615, Laboratoire des Multimatériaux et Interfaces, F-69622 Villeurbanne, France;

[2] Univ Lyon, Université Claude Bernard Lyon 1, Faculté d'Odontologie, 69008 Lyon, France

[3] Hospices Civils de Lyon, Service d'Odontologie, 69007 Lyon, France.

[4] Univ Lyon, INSA-Lyon, CNRS UMR5259, LaMCoS, F-69621, France

**Abstract:** The periodontal biomechanical environment is very difficult to investigate. By the complex geometry and composition of the periodontal ligament, its mechanical behavior is very dependent on the type of loading (compressive vs. tensile loading; static vs. cyclic loading; uniaxial vs. multiaxial) and the location around the root (cervical, middle, or apical). These different aspects of the periodontal ligament make it difficult to develop a functional biomaterial to treat periodontal attachment due to periodontal diseases. This review aims to describe the structural and biomechanical properties of the periodontal ligament. Particular importance is placed in the close interrelationship that exists between structure and biomechanics: the periodontal ligament structural organization is specific to its biomechanical environment, and its biomechanical properties are specific to its structural arrangement. This balance between structure and biomechanics can be explained by a mechanosensitive periodontal cellular activity. These specifications have to be considered in the further tissue engineering strategies for the development of an efficient biomaterial for periodontal tissues regeneration.

## 1. Introduction

Dental loss due to periodontal diseases remains a major issue for clinical dentistry. With more than 740 million cases in 2010, severe periodontitis was one of the most prevalent pathologies in human society [1,2]. Periodontitis is due to an inflammatory response initiated by an unbalanced microbiota in the oral cavity, following the deposition of the dental plaque and the accumulation of pathogenic microbial biofilm [3,4]. Such a defensive strategy from the organism may result in the destruction of the periodontal tissues [5–7], altering their biomechanical response [8,9], and resulting in the tooth loss in the severe cases left untreated.

One of the challenges of ongoing development of periodontal therapies is to achieve the repair or the regeneration of the tissues lost during periodontitis. In this context, the clinical community is increasingly focused on the use of biomaterials and particularly on tissue engineering strategy. According to the Society for Biomaterials, the aim of such implantable materials is "*to take a form*

*which can direct, through interactions with living systems, the course of any therapeutic or diagnostic procedure*" [10]. From this statement, it is clear that is necessary to accurately understand the properties and function of the living tissue in order to develop an ideal biomaterial, having specific mechanical and structural properties offering the right mechanical behavior under physiological loading, thus promoting the interactions with the surrounding cells. More specifically, tissue engineering consists in developing an engineered construct, involving a combination of cells population and biological signals, that is able to insure the injured organ's function and assist it in its regeneration once implanted [11]. In the context of periodontal disease, clinically, a biomaterial is inserted in the periodontal pocket, where the periodontal attachment is lost, in order to promote a new attachment (Figure 1, left). Guided tissue regeneration has been extensively used to heal damaged periodontal ligament. It consists in preventing epithelial migration in the periodontal pocket and thus maintaining a minimal space for tissue regeneration [12]. Nevertheless, such a healing strategy alone fails to predict the regeneration of periodontal tissues and should be combined with other regenerative approaches, such as bone graft [13]. The use of bone grafts may be limited to the regeneration of alveolar bone and does not allow a complete regeneration of the whole periodontium tissues [14]. In this context, natural polymers, such as collagen or chitosan, and artificial polymers, as Poly(lactic-co-glycolic) Acid (PLGA) or polycaprolactone (PCL) have shown interesting properties for a use as periodontal cell substrate and for medicine delivery [15]. Nevertheless, such biomaterials still fail to provide a suitable template for the formation of location-dependent guided collagen fibres, as it is the case within PDL's architecture. To insure this fibre guidance property, the biomaterial must have the suitable architecture at the time of implantation by the clinician.

In that context, tissue engineering can be considered as an interesting strategy allowing for the investigation of the biomaterial behavior in a biologically active environment seeded with different type of cells, and for the development of a biomaterial with the wanted structure prior to implantation. A key factor for the development of functional medical device through tissue engineering is the application of the suitable stimuli in order to reproduce, as faithfully as possible, the environment in which the biomaterial will be implanted to achieve its healing function. Among different type of signals, cells involved in periodontal biology are sensitive to biomechanical stimuli [16–20]. It is assumed that cells can sense a change in the interstitial fluid flow, or the stretching of its substrate, for example through a change in their cytoskeleton organization. This stimulus then influences the biochemical response of the cellular network [21,22]. Nevertheless, in the field of tissue engineering for periodontal regeneration, work has been intensively focused on biochemical cues and molecular signaling at the expense of biomechanical properties [14], while the strains sensed at the cellular level depend on the biomechanical properties of the scaffold and on the biomechanical loadings.

Periodontal biomechanical environment is very complex. If each of these components has specific mechanical properties [23–26], the cementum – periodontal ligament (PDL) – alveolar bone complex has to be considered as a whole when submitted to a mechanical loading. Although, the biomechanical response the periodontium is mainly driven by the PDL mechanical properties by absorbing the masticatory forces [27,28]. In order to maintain the homeostasis and allow the adaptation of the periodontal complex during mechanical loading, the PDL has to transfer the mechanical forces to the cementum and alveolar bone. In most of the soft to mineral tissues interfaces within the organism, this stress transfer from PDL to the mineralized tissues has to be achieved through a graded evolution of the materials properties [29,30]. Indeed, if the change in mechanical property at an interface is too steep, this might involve the development of stress concentrations, resulting in the failure at the interface [31].

In addition, and because of its specific geometry, the periodontal complex displacement field is difficult to investigate. If a lot of studies evaluate the effect of orthodontic tooth movement on the remodeling of the periodontal tissues [32,33], the displacements implied during physiological loading (e.g. mastication) or pathological loading (e.g. bruxism) are much less investigated, even if more relevant to understand periodontal regeneration.

In the view of tissue engineering for periodontal regeneration, one should wonder how the mechanical stimuli are transmitted to the cells within the whole periodontal complex [13]. This review aims to highlight the importance of the synergistic effect of the structural and biomechanical properties of the periodontal complex. This structure – biomechanics relationships can be explained by mechanotransduction and cells behavior. A clear understanding of these cellular mechanisms is of great interest to specify the factors required for the development of innovative biomaterials for periodontal tissue regeneration.

## 2. Periodontal ligament composition and structure

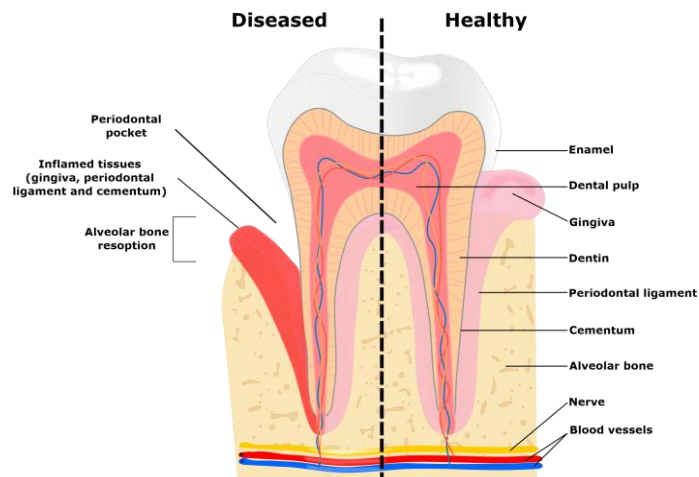

*Figure 1 Schematic overview of a healthy tooth (right) and a pathological tooth (left).*

The periodontium is the group of tissues involved in the attachment of the tooth to the dental arches. It consists in the periodontal ligament that connects the cement, coated on the dentin, to the surrounding alveolar bone (Figure 1).

Figure 2 shows light microscopy images obtained on a mouse PDL section at different locations (horizontal, oblique, apical), and can illustrate the following description. The periodontal ligament is a connective tissue made of collagen bundles immersed in a blend of water, cells, blood vessels, elastic fibres, and other proteins [34–36]. The width of the PDL reflects its high adaptability toward several types of loading, as it is indeed maintained, even if squeezed between two hard tissues, through a balance between bone formation and resorption [37,38]. The normal human ligament width ranges from 150 to 400 µm [36].

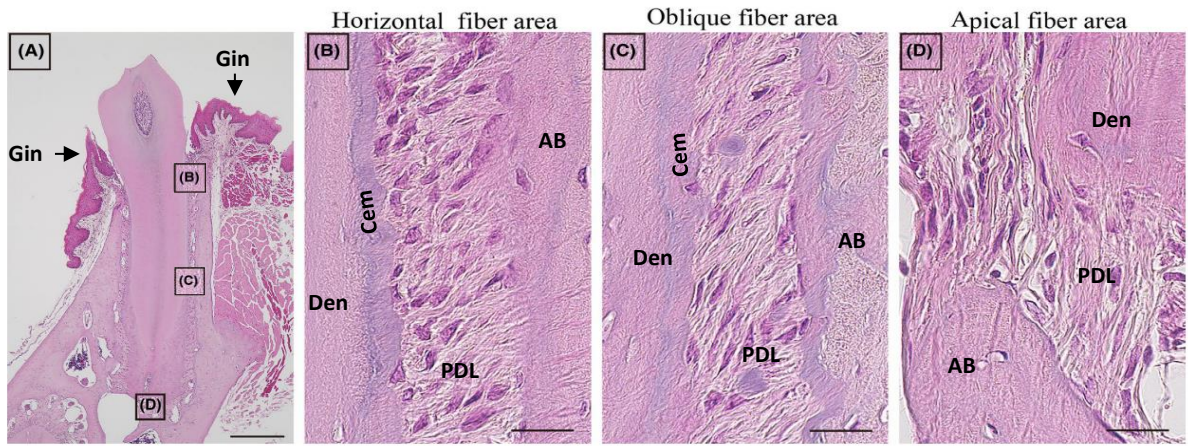

*Figure 2 Light microscopy images of a mouse PDL section observed at three locations around the dental root. A: Description of the different observation areas. B: Horizontal fibres area. C: Oblique fibres area. D: Apical fibres. Scale bars = 300 µm for A and 20 µm for B, C, and D. Gin: Gingiva, Den: Dentin, Cem: Cementum, AB: Alveolar Bone, PDL: Periodontal Ligament. From [39].*

The main types of collagen found in the PDL are type I and III, but other can be found [40–42]. Collagen bundles are composed of several collagen fibrils having the same orientation and embedded in a gel of ground substance [35,43]. These bundles are of varying thicknesses and orientations depending on their location within the PDL. Bundles with a thickness of 1-5 µm emerge from the cement, split in several smaller bundles, and gather close to the alveolar bone, forming 10-20 µm thick bundles [35,44]. The orientation of these bundles depends on their location around the dental root. At the cervical part of the root, bundles are orientated horizontally, whereas at the apical root, a vertical orientation is observed. In the middle part of the root, which represents the largest region, the collagen fibres are orientated obliquely, from the root up toward the alveolar bone (Figure 2) [39]. In the majority of cases, the investigations on the organization of collagen fibres in the PDL have been conducted on mandibular teeth, some results suggest that in the maxillary teeth, the oblique PDL might be oriented similarly, from the bone down to the cement. In that way, in the maxillae, the fibres are oriented in a manner to preferentially resist to an extrusive force compare to an intrusive force [45]. In reality, the organization of the collagen fibres between the cementum and the bone might be much more complicated, forming a connected 3D network [39,46,47]. At the interfaces with the cementum or the alveolar bone, wide collagen bundles, called Sharpey's fibres, are embedded within the mineralized matrix [48,49]. This attachment is believed to be assisted by the presence of circumferential collagen fibres around the tooth and the bone [46]. Generally, in human soft to mineralized tissues interfaces, the collagen of the soft tissue is gradually mineralized as it gets closer to the interface [50]. Despite the difficulty to investigate accurately this area due to its small geometry, it is believed that such a gradient in material properties may exist at the PDL entheses [46].

The interstitial space between the collagen fibres and bundles is filled with various types of components making a gel called ground substance. It has been measured that, within a collagen bundles, the ground substance fill 65 % of the volume, the resting 35 % being filled with the collagen fibrils [35]. Moreover, the collagen has been found to be in a lower quantity in the apical region of the tooth, compare to the cervical or the middle parts [35,51].

In order to insure a necessary supply in nutrients, the PDL is crossed by blood vessels, with a higher density at the cervical part of the root compare to apical region [52]. These blood vessels are ranged from 5 to 50 µm in diameter [53] with a main orientation along the occluso-apical direction [54].

Oxytalan fibres are observed closely associated with the vascular network [55]. Oxytalan is an elastic fibre that is thought to play a role in the preservation of the vascular vessels and the regulation of blood flow depending on the tooth function [56–58]. They are approximately 60 µm in diameter [59].

Non-collagenous proteins, such as proteoglycans (PGs), osteopontin (OPN), or alkaline phosphatase (ALP) are observed within the PDL tissue [38,60,61]. One function for these proteins is associated with the regularization of mineral formation within the extracellular matrix. It has been observed that the amount of such proteins is more important at the interface between the PDL and the two mineralized tissues [46,62].

The PDL is known to be a source of cells for its surrounding environment. The periodontal ligament stem cells (PDLSCs) can differentiate into fibroblasts, osteoblasts, or cementoblasts to synthesize the ligament, the alveolar bone, or the cement, respectively [63–67]. PDL cells have been regarded as the cell type with the highest potential for periodontal regeneration [67]. These cells are often observed close to the blood vessels [68]. Within the PDL, the PDLSCs mainly differentiate into fibroblasts that remove the old collagen and synthesize the new one [69]. In certain regions of the root, the fibroblasts have the same orientation as the collagen bundles and form a connected network around them [39,51,70]. Whereas at the interface between the PDL and the mineralized tissues, the PDLSCs differentiate into osteoblasts or cementoblasts [71]. It has been shown that PDL fibroblasts can also differentiate into osteoblastic cells [65,72]. It is proposed that the cells of the PDL, the cement, and the alveolar bone are all connected, forming a unique cellular network, similar to the lacuna-canalicular network observed in bone tissue, promoting the exchanges between them [51,73,74].The remaining of the non-collagenous part of the PDL is mainly built on water and blood that fill of the space [36].

This whole complex architecture is developed to provide specific biomechanical properties.

## 3. Periodontium biomechanics

### 3.1. The Cementum – PDL – Alveolar bone complex: global mechanical response

Because of its role within the organism, the cementum – PDL – alveolar bone complex has to show both strong and soft mechanical properties. The complex has to be stiff enough to allow the mastication of hard diet product. In the other hand, the load transfer between the tooth and the alveolar bone should be low enough in order to avoid any tissues damage. In 2019, Ben-Zvi et al., obtained the force-displacement curve measured by loading of a pig molar not extracted from its jaw, to simulate an *in vivo* loading (Figure 3b) [75]. This curve shows an increasing stiffness of the periodontal complex with increasing load: at higher loads, the complex is more difficult to deform. An ideal medical device for periodontal regeneration should undergo the same mechanical response when implanted within the oral cavity. *In vivo* measurements are more difficult to perform and assess. In 2016, Keilig et al. obtained the force displacement curves of human incisors loaded in the horizontal direction (labio-oral) at different displacement rates [76]. Even if this type of loading does not represent a typical clinical situation, it has been chosen because of the limited space in the oral cavity (Figure 3a). This investigation showed that the mechanical response of the cementum – PDL – alveolar bone complex depends on the loading rate. At higher rates (0.1 s in Figure 3a) the stiffness of the complex appears to be constant, whereas at lower rate, it varies with the magnitude of loading.

During mastication, the mechanical response of the cementum – PDL – alveolar bone complex is mainly attributable to the PDL mechanical behavior [27,28]. The PDL is softer than the cementum or the alveolar bone: it undergoes higher strains when the complex is subjected to an external loading. By basically considering the periodontal complex as the succession of the cement, the PDL and the alveolar bone in series, each of them being a spring with the stiffness $k_{Cement}$, $k_{PDL}$, and $k_{Alveolar\ Bone}$ (N.mm$^{-1}$), respectively, the equivalent stiffness of the periodontal complex can be measured as follows (Eq. (1)):

$$\frac{1}{k_{eq}} = \frac{1}{k_{Cement}} + \frac{1}{k_{PDL}} + \frac{1}{k_{Alveolar\ Bone}} \qquad \text{Eq. (1)}$$

As the cementum and bone are much stiffer than the PDL [46], $k_{Cementum} \gg k_{PDL}$, and $k_{Alveolar\ Bone} \gg k_{PDL}$, signifying that the apparent stiffness of the periodontium, $k_{eq}$, is almost the same as the PDL's one.

One major difficulty in understanding periodontium biomechanics is that it undergoes various types of loading, from physiological ones (such as masticatory forces), to pathological ones (such as bruxism), including therapeutic loadings (such as orthodontic forces). If it is assumed that the loss of periodontal attachment initiates through a change in its loading environment, the influence of such abnormal loadings on a pathological periodontium is less stated [77,78]. These different types of loading differ in magnitudes, frequencies, and orientations. If orthodontic movements involve low static forces ($10^0$ N, [79]), masticatory forces are much higher loads ($10^2$ N, [80–82]) applied cyclically [83], and can be even higher in the case of abnormal occlusion [84]. When a static loading is applied on the tooth, the PDL undergoes a creep behavior: it first reacts by counteracting this load, and then adapts to this load through a microstructural re-organization in order to decrease its state of stress [85–88]. At the opposite, during a cyclic loading, the periodontal complex undergoes a hysteretic behavior, signifying that the response under loading differs from the response during unloading [89–91]. Moreover, the application of a same load on a healthy and injured periodontium will not result in the same strain and stress distribution due to the different geometries [8]. As it can thus be understand, the periodontal complex cannot be modeled by a simple spring characterized by a unique simple stiffness ($k_{eq}$ (N.mm$^{-1}$)) but as a combination of springs and dashpots undergoing a complex mechanical behavior depending on the PDL geometry and on the type of loading [92].

The understanding of stresses and strains distribution around the dental root has been increasingly improved with the development of numerical approaches [93–95]. These tools allow the consideration of the complex geometry of the root, that is involved in the mechanical response of the periodontal complex [27,96–98], and considering various complex loading conditions that are difficult to achieved in experimental studies [95,99]. For example, these models highlighted the significance of the fluid phase in the viscous behavior of the PDL under a physiological loading [100–102] and other different microstructural features [103,104].

Despite this great interest, numerical modeling fails in providing a clear statement regarding PDL mechanical response [92]. The limitation arising from these numerical studies may be explained by the complex PDL's microstructure and the contribution of components in its mechanical behavior. The local property of the tissue depends on its local structure, and evolves depending on the location around the dental root. It is thus important to know the structural and associated biomechanical properties of the PDL in order to understand the specific distribution of stress and strain in the PDL and transmitted to the cementum and to the alveolar bone.

## 3.2. PDL's biomechanical properties

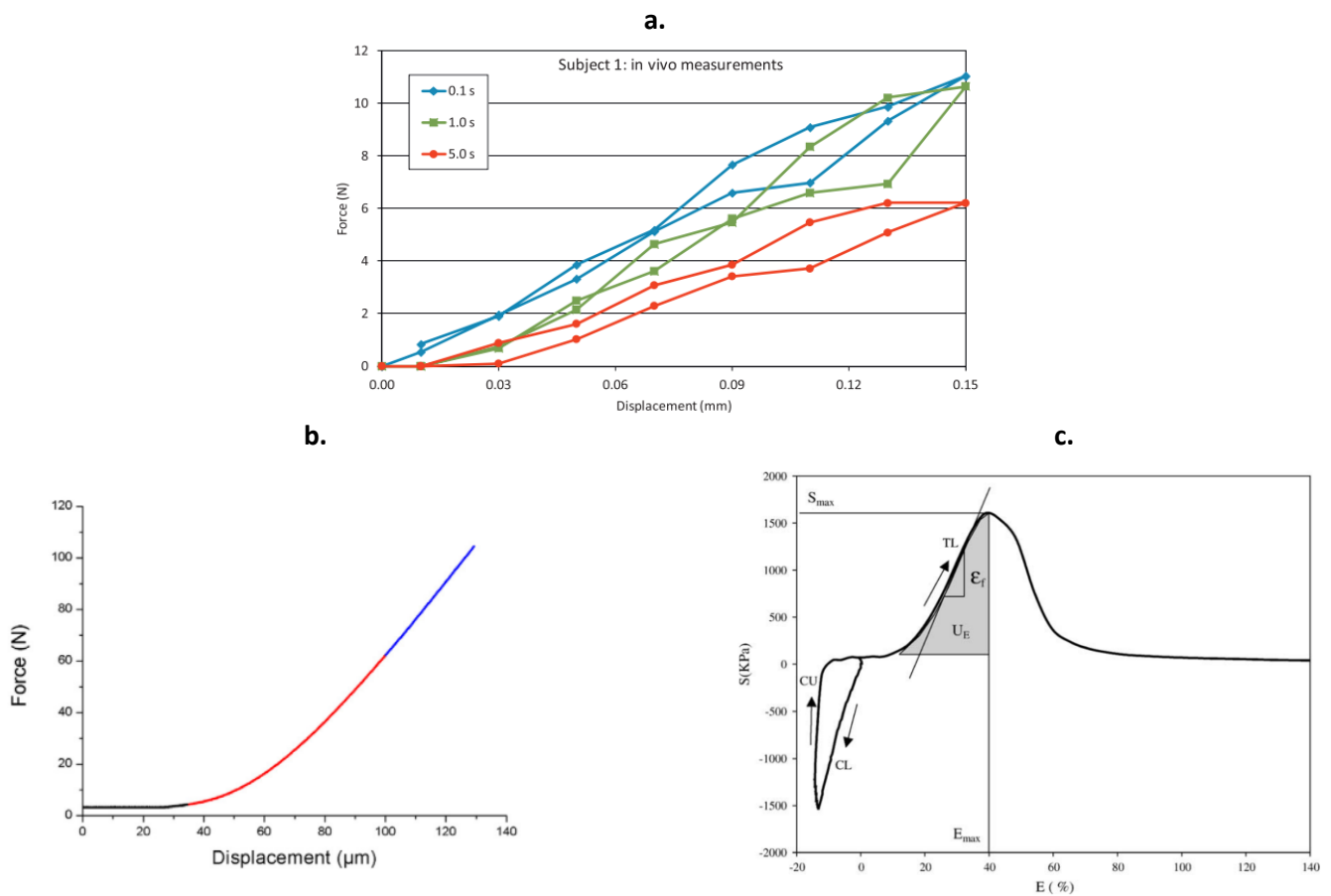

Figure 3 a: Force displacement curves obtained in vivo on human incisors in the labio-oral direction at different displacement rate (0.15 mm in 0.1 s, 1.0 s or 5.0 s. From [76]). b: Force displacement curve from a mini-pig not extracted molar loaded in axial compression, thus simulated an in vivo condition (from [75]). c: Stress-strain curve of an extracted bovine molar in compression (E (%) < 0) and in tension (E (%) > 0). The protocol included compressive loading (CL), unloading (CU) and tensile loading (TL) up to failure. The analyzed mechanical parameters were elastic modulus ($\varepsilon_f$), maximal stress ($S_{max}$), maximizer strain ($E_{max}$) and strain energy dens ity ($U_E$)  (from [105]).

PDL's biomechanical properties are the utmost importance as it acts as a damper to the high loads applied by the teeth and their periodontal supporting tissues during mastication. Because of the complex geometry of the tooth, the periodontal ligament is subjected to different types of strain depending on the location under the tooth, with a maximum tensile strain on the side of the root and a maximum compressive strains on the apexes when the tooth is exposed to an external loading [94]. The orientation and morphological properties of collagen bundles have been shown to be correlated with the type of loading they handle [51]. This is why it is important to understand the behavior of the PDL under various types of loading.

The general behavior of a bovine PDL under both tension (E (%) > 0) and compression (E (%) < 0) can be seen in Figure 3c [105]. Regarding Figure 3, it is interesting to see that, considering the lower displacement rates for the *in vivo* experiment (1.0 s and 5.0 s in Figure 3a), we can observe for the three different types of experiments (Figure 3a, b, and c) an evolution of the stiffness of the tested material, with almost no change in the force for small displacements, even if this evolution is more pronounced for the *in vitro* experiments (Figure 3b and c).

Under tension, the PDL shows hyperplastic properties, meaning that its stiffness evolves as the loading increases [106]. This behavior under tension is mostly due to an evolution in the organization of the collagen bundles. In a steady state, collagen fibres of the periodontal ligament are entangled. The first non-linear region of the stress-strain curve (Figure 3c, E (%) > 0) might be due the disentanglement of these bundles [107]. While the collagen is organizing in fibres orientated along the loading direction, some fibres start to be stretched resulting in an increase the apparent elasticity of the tissue. When all the fibres are stretched, the behavior of the PDL is linear until the progressive failure of the collagen bundles [108]. The mechanical strength of human periodontal ligament under tensile testing has been measured to be around 3 MPa [109].

As the PDL undergoes a non-linear behavior under tension, its elastic modulus evolves while the load increases. Considering the final linear part of the curve, it has been observed that the elastic modulus of the PDL depends on the location around the root and ranged between 1 and 10 MPa on bovine PDL or around 5 MPa on pigs for the high stresses area (as shown by $\varepsilon_f$ in Figure 3c) [90,105]. The low modulus has been measured around 0.15 MPa (corresponding to the low deformations (0 < E(%) < 10) in Figure 3c) [90]. No real trend about the dependence of this tensile elastic modulus with loading rate has been measured by Nishihira et al. and Bergomi et al. [110,111]. Under static loading in tension, PDL undergoes a creep behavior, signifying that, at a constant stress, the PDL still deforms with time, possibly due to a structural rearrangement of the collagen network [112]. Similarly, the PDL undergoes stress relaxation, meaning that at a constant deformation, stress decreases within the PDL [112].

At the fibril scale, the elastic tensile modulus of collagen has been measured around 100 and 500 MPa on fibril $10^2$ nm in diameter using micro tensile experiments [113,114]. A modulus ranged from 1 to 10 GPa has been measured by nanoindentation, providing information in the transverse direction [115]. A viscous component of the PDL in tension can be associated with the viscous component of the collagen itself. More specifically, collagen undergoes creep behavior during static loading [116,117]. This mechanism operates through a reorganization at the molecular scale [118].

It has also been observed that during loading, some singularities occurred at the interface between the PDL and the cementum or the alveolar bone [108]. The apparent stiffness of the PDL increases as the collagen fibres become embedded in the mineralized tissues at the entheses [46]. At these interfaces, the variation in stiffness is at the origin of stresses and strains gradients [119]. More specifically, in a graded material, strain is higher in the softer part [120]. Bone remodeling has been shown to be increased in areas of high strain gradient, resulting in the formation of a material with increased stiffness in these areas [121]. Moreover, higher loading state is associated with thicker collagen bundles [51]. These observations are in accordance with the graded increase in mineral content and the higher thickness of Sharpey's fibres at the interface between the PDL and the cementum or the alveolar bone [35,46]. It has been observed that the elastic modulus of single collagen bundle increases with the increasing content of mineral [122]. Moreover, the increased number of collagen fibrils in thicker bundles increases its apparent stiffness by multiplying the number of springs (collagen fibrils) associated in parallel.

Under compression, the PDL shows a viscous behavior, meaning that the relationship between strain and stress during loading is different to the one during unloading. This is represented by the hysteresis loop on Figure 3c (E (%) < 0) [123]. Moreover, under compression, PDL's mechanical response depends on the loading rate, with a higher maximum compressive stress with increasing loading rate [111]. This particular compressive behavior is believed to be associated with the interaction of fluid and solid phases during loading [111,124]. As a hyperelastic tissue, a low compressive elastic modulus of about 0.5 MPa and a high modulus of 20 MPa have been measured on cat and dogs PDLs [110].

In reality, the PDL's mechanical behavior during compressive loading depends on periodontium components' permeability that has been estimated around $10^{-14} m^4/N.s$ and $10^{-17} m^4/N.s$ for alveolar bone and cementum, respectively [100,124]. Indeed, as the local volume under compression decreases, the diffusion of water and other ground substances directly depends on the materials' permeability and its ability to let the fluids flow to another less constrained region. It is believed that with increasing loading magnitude, the collapse of the blood vessels network results in a decrease in permeability, thus explain the non-linear mechanical response of the PDL under compression [125]. It has been shown that the amount of ground substance and water is higher in region subjected to compressive loading, such as the apical part of the root [35]. This highlights the importance of water and other fluids in the mechanical behavior of the PDL under compression. The rate dependence of PDL's mechanical response under compression may also be explained considering the flow of interstitial fluids within PDL's structure. At low rate, these fluids have the time to move to another region [126,127]. At increasing rates, the ground substances might be squeezed to fast, and its compressive behavior is added to the whole PDL behavior. This is of great importance since chewing activity implies loading frequencies ranged from 1 to 1.5 Hz [83]. The viscous behavior of PDL during the unloading part of the compressive test (Figure 3c, E (%) < 0) is partly due to the refilling of the interstitial spaces by the ground substance. This fluids exchange has been shown to be enhanced at the alveolar bone-PDL interface, and almost insignificant at the cement-PDL interface, due to the low permeability of the cementum compare to alveolar bone [100].

As can be seen with this description, the PDL's biomechanical environment is very complex. Data regarding the distribution of stresses and strains around the PDL would be of great interest to better understand this biomechanical environment. A better knowledge on the local mechanical properties around the PDL may be needed to improve the numerical models developed to estimate the strains and stresses fields. For example, nanoindentation experiments could be performed on different location to evaluate the variation of properties around the root [128]. *In vivo* stresses and strains distributions would allow to better associate the external mechanical loading applied on the tooth and the tissue produced by the cells.

# 4. PDL mechanotransduction or how to explain the relationship between structure and biomechanics

Mechanotransduction is the mechanism by which the cellular activity is linked to its biomechanical environment. It is known that some of cells' activities are driven by their strain state [22]. The cell's biomechanical environment has an influence on its strain state under an applied external loading, by inducing cellular deformation through the direct deformation of the PDL's extracellular matrix or by the fluid flow induced shear strain in both the PDL and the bone [129].

It is known that cells orientation depends on their substrate deformation in two dimensions [130]. More specifically, cells' orientation has been shown to be parallel to the principal orientation of deformation under static loading and perpendicular to it under cycling loading on a 2D stretched membrane [131,132]. Similarly, a flow induced shear strain promotes the alignment of endothelial cells in the direction of flow [133]. The orientation and cell's activities also depends on the amplitude of the deformation [134,135]. A cyclic mechanical loading applied parallelly to the main orientation of a nanofibrous poly-caprolactone/gelatin scaffold has been shown to promote the PDL's cells alignment along the fibres whereas on a randomly oriented scaffold, the cells were oriented perpendicular to the applied loading [136]. This phenomenon is known as contact guidance [137]. Furthermore, it is known that cells are sensitive to the substrate stiffness [138]. In three dimensions, static and dynamic mechanical loading induced the alignment of dermal fibroblast parallel to the direction of loading cultured in a fibrous media [139]. Moreover, unidirectionally loaded tendon fibroblasts have been shown to produce a highly anisotropic collagen fibrous media whereas bi-axially loading promotes the development of a randomly arranged collagen fibres [140]. Cell's and collagen degree of alignment can thus provide relevant data on the PDL structure – biomechanics interrelationship.

It has been observed that periodontal cells' orientation varies depending on their location around the dental root, with almost perpendicular, parallel, and randomly oriented cells in regard to the collagen bundles on the cervical, the middle, and the apical part of the root, respectively (Figure 4) [141]. These results imply that, depending on the location around the root, the periodontal cells do not undergo the same amplitude and type o strains (static or cyclic, uniaxial or multiaxial), and thus do not have the same activity. For example, it might be interesting to investigate if the alignment between periodontal cells and collagen bundles in the middle part of the dental root may arise from static occlusal forces applied during daily activity [142], or even from gravity that can have an influence on the zero strain state of soft tissues [143]. As enounced previously, at the oblique locations, the PDL fibres spread from the bone to the cementum in a cervical-apical orientation (see Figure 2C). But if the majorities of these investigations have been performed on the mandible, some results suggest that the PDL fibres in the maxilla are oriented from the bone down to the cementum, in an apical-cervical orientation, as if the fibres prevented the tooth from falling from the maxillary bone [45]. But to the authors' knowledge, a direct comparison between mandible and maxilla PDL organization has never been performed. Moreover, while numerous studies having shown a significant influence of intrusive or extrusive loadings, considered forces in the order of the force of gravity on tooth mobility [144–148], the influence of gravity forces alone is not known. According to the authors, this is an interesting point to consider, because gravity is associated to a static loading that might have an influence on the specific orientation at this location around the PDL. This highlights the importance for understanding the influence of the different type strains on the cellular activity within the PDL.

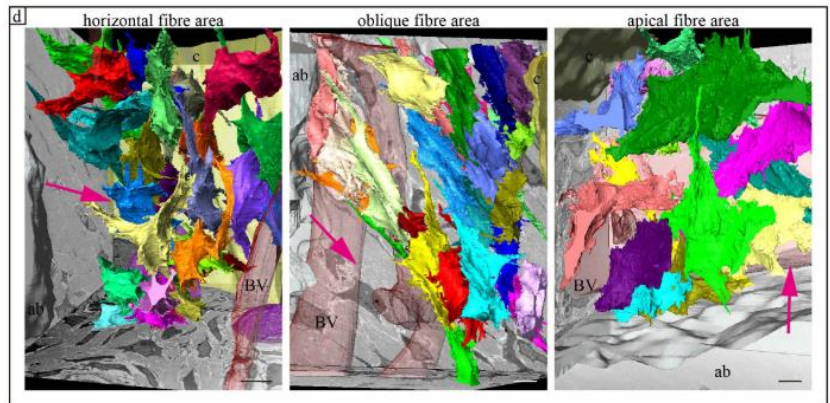

*Figure 4 Focused ionized beam / scanning electronic microscope 3D reconstructed images of PDL cells in horizontal, oblique, and apical regions of the root, on a mouse model. The arrows indicate the collagen bundles direction. ab= alveolar bone, BV = blood vessel. Scale bars = 10 µm (from*[141] *).*

A tensile strain is perceived by the cells when the substrate on which they adhere is stretched. It has been shown that periodontal fibroblasts are sensitive to stretching, up to 9 %, by adapting their cytoskeleton along the direction of the stress [149–152]. It is assumed that this stretch sensitive property is due to the binding of cells to the extracellular matrix through integrins that are proteins with an intra- and an extracellular part [153,154]. In response to slight cyclic stretching, periodontal fibroblasts produce a higher quantity type I collagen extracellular matrix compared to unstretched control fibroblasts. However, stretching at magnitude higher than 10 % resulted in no difference between experimental and control [19]. In a same way, tensile cycles are believed to up-regulate the production of metalloproteinase proteins needed in the degradation of old or damaged collagen at low (1.5 %, [155]) or high (10 %, [156,157]) magnitude of stretching. Tensile strain also plays a role in the bone remodeling at the interface between the bone and the PDL. A cyclic tensile loading has been shown to up-regulate the expression of osteogenic genes [158] while inhibiting osteoclastogenesis [159]. Moreover, the strain gradient occurring at the entheses might be one explanation of the graded increase of mineral content at the interface between mineralized tissues and PDL [119].

The compressive strains submitted at the cellular level are due to hydrostatic pressure exerted by the interstitial fluids. As for the tensile loading, the compressive strains are sensed through an increased attachment of the integrins to their substrate [40]. The synthesis of type I collagen by periodontal ligament fibroblasts is not up regulated when subjected to a compressive cyclic loading whereas type VI collagen has been shown to be produced [40]. Collagen type VI is known to mediate cell attachment [160]. It is well assumed that static compression loadings involved in orthodontic movement of are associated with the resorption of the alveolar bone [161]. Static compressive load applied to fibroblasts indeed resulted in the gene expressions related to bone resorption [162–166].

In the case of shear loading, stress and strain develop because the applied mechanical loading pushes the cell to move whereas the adhesive substrate applies a resistance to retain it. As we have seen previously, under compression, the PDL's behavior is driven by the flow of interstitial fluids [124]. These fluids flow, not perfectly perpendicular to the substrate, result in the development of shear strain and stress at the cellular level. Under pulsating shear flow, an increase in nitric oxide and Prostaglandin E2 production by periodontal fibroblast has been measured [16]. These two molecules are involved in collagen turnover [167,168]. Moreover, static fluid shear stress is assumed to be involved in the differentiation of PDL cells into osteogenic cells [17,169].

Regarding the mechanobiology of the alveolar bone, the major role of the bone embedded osteocytes in orchestrating the bone remodeling process, through bone osteoblasts and osteoclasts supply, is well stated [170,171]. The bone being much stiffer than the PDL, the deformations underwent by the tissue are very small. The bone remodeling process is thus mainly driven by a flow induced shear strain on the membrane of the osteocytes [172]. Within bone tissue, the osteocytes are all connected through a set of canaliculi going from an osteocyte to another. These canaliculi, 250 nm in diameter [173], are filled with interstitial fluids [174]. When the tissue is strained, the flow of these fluids increases and the shear stress on the osteocytes membrane is modified [175].

The cementum is a particular tissue as it contained cells only on the apical part of the root. This implies that on the middle and cervical region of the root, the turnover potential for the acellular cementum is limited. Thus, the precise cementum remodeling mechanisms are not well-known. It is believed that the remodeling of cementum might be similar to the bone remodeling process. Indeed, a cementocytes network, similar to the osteocytes network in bone, is formed within the cementum [176] and similar microstructures with slight differences can be observed [177].

The bone and cementum remodeling mechanisms highlight the importance of the entheses biomechanics and their ability to transfer loads from the PDL to modify the fluid flows within these mineralized tissues [29,94,178].

# 5. Tissue engineering for periodontal tissue regeneration: toward the development of an ideal biomaterial

## 5.1. Current biomaterials: current status and future perspective

Currently developed biomaterials for periodontal regeneration, including monophasic or composite materials, aim to provide the most suitable biochemical and biomechanical microenvironment for cells adhesion and tissue production [15]. The biochemical pathway is generally achieved by incorporating biological agents involved in the PDL's cells activity. The biomechanical properties of the biomaterial have been much less studied. They can be tuned by the association between the composition and the 3D architecture of the scaffold. The difficulty in this is to find the suitable compromise between the apparent mechanical properties needed to support the whole structure, and the material's mechanical properties that are directly related to the mechanical forces transmitted at the cellular level.

As natural polymers, collagen, gelatin, or chitosan based biomaterials have been extensively used to provide a good substrate for PDL's cells culture [13,15,179]. In addition to their high biocompatibility, they have been shown to promote adhesion and growth of periodontal fibroblasts [179–181]. Loaded with biological cues, they also show interesting properties to regenerate periodontal mineralized tissues [182,183]. Thin films of such natural polymers have an apparent elastic modulus around 100 MPa [184]. As stated before, the apparent modulus of the PDL has been measured from 0.15 to 10 MPa (see section 3.2). In this bulk form, such natural polymers-based biomaterials are thus not suitable for periodontal ligament regeneration application. The apparent elastic modulus of collagen constructs can nevertheless be tuned by modifying their internal organization. By arranging collagen fibres into bundles (as it can be observed within the PDL), one can decrease the collagen apparent modulus from 180 MPa (fibre) to 27 MPa (bundles) [185]. This can be of great interest for periodontal regeneration application.

Some artificial polymer-based biomaterials have shown interesting properties regarding periodontal regeneration. Poly(lactic-co-glycolic) Acid (PLGA) is a biodegradable co-polymer that has been extensively used for biomedical applications [186,187]. Similarly, polycaprolactone (PCL) has also been used for periodontal regeneration [188]. Foams made of PLGA or PCL have an elastic modulus around 5 MPa, which is similar to the PDL's apparent modulus [189,190]. At the fibre's scale, PCL shows an elastic modulus close to 100 MPa measured by micro tensile experiments and on the order to the GPa using indentation techniques [191,192]. With relatively similar properties, PLGA has a modulus close to 100 MPa and 5 GPa measured by micro tensile or indentation tests, respectively [193,194]. This distinction between apparent and material's scale elastic modulus is of great importance as the apparent modulus will drive the apparent behavior of the construct under a specific loading, whereas the material's scale modulus will drive the cellular local behavior at the surface of these substrates [138]. Furthermore, both PLGA and PCL undergo creep behavior under static loading [195,196]. These mechanical properties at different length scale are close to the PDL's ones, that makes PCL and PLGA as promising scaffold candidates for periodontal ligament regeneration strategies.

The addition of bioactive materials in these polymer scaffolds, such as hydroxyapatite or tricalcium phosphate, also appears to promote bone or cementum gene expression by periodontal cells [197,198] and to increase their elastic modulus or strength [199,200]. Mechanical properties can also be tuned by providing the scaffold a specific architecture, for example by using electrospinning or other additive manufacturing techniques [190,201–203].

In order to promote the regeneration of the whole cementum – PDL – alveolar bone complex, numerous studies have proposed multiphasic scaffolds that mimics the periodontium architecture [204–207]. These types of scaffold are mainly made of polymeric scaffold between two stiffer and bioactive layers. But in such studies, no attention is paid to the permeability of the external layers.

Despite all of these major efforts, little attention has been paid to the state of strain underwent by the cells within the scaffold whereas, in the authors' opinion, it is a major factor to consider for the development an efficient biomaterial for periodontal ligament regeneration. Once implanted in the periodontal defect, the cells seeded within the scaffold will be subjected to a complex biomechanical environment. The scaffold, depending on its apparent mechanical properties, will deform according to the external loading and the geometry of the tooth. These deformations will be transmitted at the cellular level depending on the true stiffness and permeability of the material, and the architecture of the scaffold. Thus, all of these properties are of major importance for the development of engineered materials for periodontal regeneration.

According to the periodontal ligament properties, the biomaterial should be anisotropic, such as an oriented fibrous scaffold. Its apparent tensile elastic modulus should be close to 150 KPa for small deformations (from 0 to 10 %) and around 5 MPa for higher deformations, in the fibres direction [24]. In the case of a multiphasic scaffold, the external layers should have a permeability in the order to $10^{-14}$ m$^4$/Ns to allow the suitable compressive and shear strain on the cells [100]. At the cellular level, the materials should have a modulus close to collagen modulus, that is 0.1 to 0.5 GPa in the longitudinal direction and between 1 and 10 GPa in the transverse direction for fibres of $10^2$ nm in diameter [113–115].

Because of the impossibility to immobilize the damaged region, as it can be done for bone fracture healing for example, an ideal biomaterial should have the suitable biomechanical and structural properties at the time it is implanted within the periodontal defect [208]. To achieve this objective, tissue engineering through bioreactor stimulation appears as a promising strategy [136,139,209]. Bioreactor can be used to stimulate biochemically and biomechanically a scaffold seeded with cells to produce a specific tissue. By adapting the mechanical loading to the type of scaffold, and by mimicking the tooth geometry in the bioreactor, one can simulate the periodontal ligament biomechanical environment. An *in situ* characterization of cellular membrane through fluorescence, can allow assessing the strain state at the cellular level by measuring the change in the shape of the cellular membrane before and after loading [210].

### 5.2. Biomechanical specifications

Regarding this state of the art, there are no more consensuses about the biomechanical specifications that are needed to be considered in the further development of a functional biomaterial for periodontal tissues regeneration. PDL's mechanical environment has an influence on collagen fibres orientation and thickness [140,211], periodontal cells orientations, morphology and activities [51,130,135], and composition of the ground substance [212]. The authors now proposed some biomechanical guidelines that should be followed for the further development of an efficient biomaterials for periodontal regeneration using a bioreactor. Figure 5 gives a summary of these specifications. Obviously, the biochemical and biological cues are other important factors that need to be precisely controlled during the manufacturing process.

An important parameter that needs to be controlled in cells mechanosensitivity is their deformation state under an external applied loading *in vitro*, for example by *in situ* observation of the cellular membrane through fluorescence microscopy [213]. The strain state depends obviously on the type of loading, but also on the elasticity and permeability of the scaffold. They will drive how mechanical forces are transmitted from the external loading device to the cellular membrane. The stiffness ability of the scaffold will define the stretching of cells whereas its permeability will drive the fluid flow on the cellular membranes. These parameters thus have to be controlled, or at least known, in order to consider them in the cellular response [214].

The periodontal ligament has a complex geometry implying a complex strain distribution depending on the location around the root [94]. In 2009, Berendsen et al., developed an experimental device allowing a cyclically load in hydrogel containing PDL fibroblasts scaffold with cylindrical geometry. Within this set up, the hydrogel and fibroblasts were subjected to shear strains. The results showed that the production of type I collagen by the PDL fibroblasts was up-regulated when the displacement applied to the cylindrical artificial root was about 200 µm [215]. This study is of great interest as it is the first that investigate geometry close to the dental root geometry. However, one can wonder if the hydrogel scaffold, a very soft material, was the most suited for periodontal regeneration model.

As we have stated previously in the current review, fibroblasts appear to be aligned to the collagen bundles in the middle part of the PDL [141]. This configuration suggests that the fibre-based architecture of the PDL has an importance in biological process associated to periodontal regeneration. It has been observed that periodontal fibroblasts naturally aligned with the principal orientation of the fibres when cultured in an electrospun scaffold, even without any applied stretching [207,216].Different technologies may help to obtain geometries similar to biological fibrillar geometries, with evolving preferential directions [217], or various fibres thickness [218].

Another important feature of PDL is its anchorage on its both sides to the cementum and the alveolar bone. This sandwiched configuration results in a non-homogeneous strain field within the PDL between the tooth and the surrounding bone due to its graded property at the entheses. The arising strain gradient at the interface is believed to be of major importance for the load transfer from the PDL to the mineralized tissue to allow their remodeling and to avoid the failure at the interface by reducing the stress concentration state between two dissimilar materials [50]. Berendsen et al. coated the walls of chamber with a calcium phosphate based materials to promote the anchorage of the scaffold and simulate the presence of entheses [215]. To our knowledge, no information about the influence of this attachment zone on the cellular activities was demonstrated. Moreover, considering the alignment between PDL cells and collagen bundles on the obliquely orientated part of the PDL collagen, this anchorage may play a role in the support of the tooth against the static physiological loading.

In order to develop a biomaterial that might be able to sustain the mechanical environment of PDL when implanted, one should consider physiological loading involved in daily activities. A recently developed experimental device allows applying realistic chewing kinematics to artificial mandibular and maxillary dental arches [219,220]. This kind of experimental devices could be used to measure the displacement kinematic of the dental root under physiological loadings (work in progress), and to validate the final mechanical properties of the biomaterials purposed for periodontal regeneration under realistic loadings.

The final step in the development of such a biomaterial is its validation in an environment close to human periodontium. The final biomaterial has to show biomechanical properties that allow its resistance during mastication [75,219]. The validation of structural properties of the PDL could be performed by comparing with reference values for collagen and periodontal cellular network as proposed by Alves et al. [221].

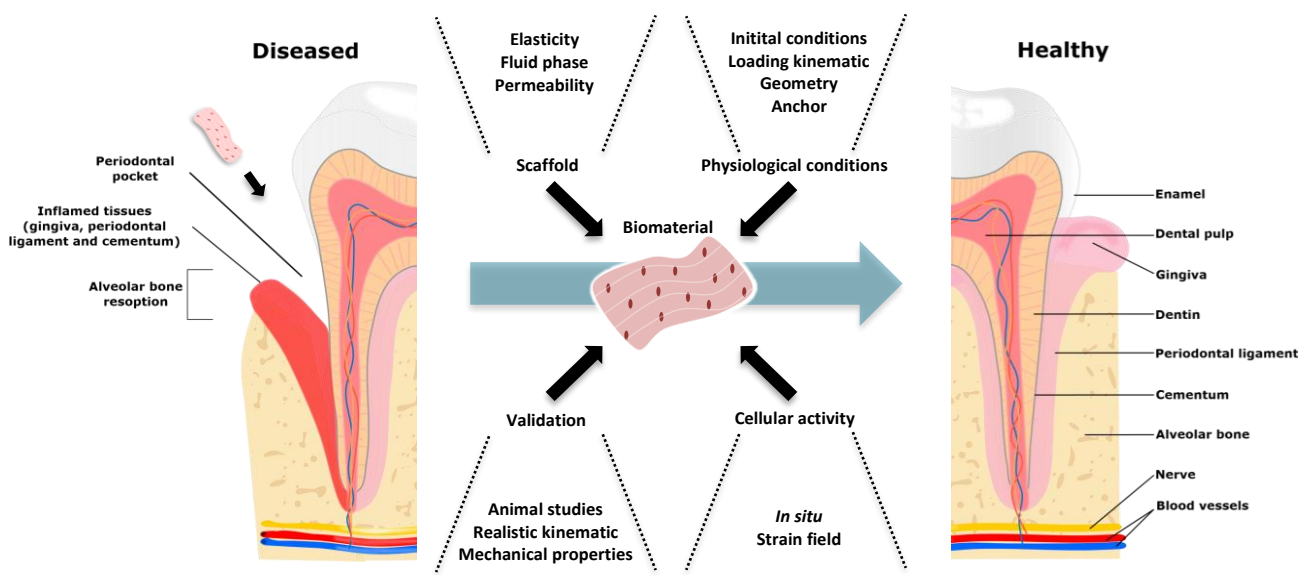

*Figure 5 Summary schematic of the biomechanical specifications that have to be considered for the development of an ideal biomaterial for periodontal tissue regeneration. Tissue engineering strategy involves cellular culture within a scaffold subjected to specific stimuli. The scaffold materials properties drive the mechanical stimuli transmitted to the cells. Stretching of the cells is driven by the elastic properties of the scaffold, whereas compressive and shear strains are driven by the presence of a fluid phase* [214]. *To promote the production of a tissue similar to the human PDL by the cells, the applied mechanical loading should be similar to physiological conditions. Particularly, the geometry and anchor have an*

*influence on the distribution of stress and strain around the scaffold. An initial condition of loading should be investigated to understand the cells orientation in the obliquely orientated fibres. Loading kinematics associated with mastication, deglutition or other physiological activities should also be investigated* [219]. *The cellular activity being related to these mechanical stimuli, an in situ observation of the strain field at the cellular membrane is of great interest to control the load transfer* [213]. *Last but not least, the properties of the final biomaterial have to be validated in order to ensure that it will provide a suitable environment for the cells when implanted in the human body. The elastic and ultimate properties of the biomaterial have to be compared to a healthy human PDL. Its mechanical fatigue should be evaluated within a realistic mechanical environment* [219]. *Finally, its behavior within a physiological environment can be evaluated using animal models* [221]. *It has not been discussed in the context of this review article, but the perfusion fluid composition and types of cell used for the development of such a biomaterial are also of major interest.*

Besides the technological challenge that is the development of an ideal biomaterial, the aim for the definition of such biomechanical specifications is to improve our scientific knowledge regarding periodontal regeneration. The best strategy to accurately control the behavior of an implanted biomaterial is to perfectly understand its biological, biochemical, and biomechanical environment: a better understanding for a better healing. By gathering experimental data and theoretical modeling, the discipline of mechanobiology appears as essential to improve our knowledge regarding biological mechanisms [222]. A few mechanobiological models have already been proposed in the past. But the models developed so far only consider mainly consider simple static loadings conditions often associated to orthodontic movements [223–225]. The biomechanical specifications defined in the current review may also be of interest for the further development of mechanobiological models to better understand the biological mechanisms involved in periodontal tissues regeneration.

# 6. Acknowledgment

This work was founded by the ANR project Toothbox (ANR 16-CE08-0024). The authors want to thank Charlène Chevalier for the drawing of the schematics.